\documentclass[9pt,lineno]{article}
\pdfoutput=1
\usepackage{float}
\usepackage[margin=0cm]{caption} 
\usepackage{subcaption} % For subfigure
\usepackage[utf8]{inputenc}
\usepackage{amsmath,amsfonts,amssymb}
\usepackage[colorlinks=true, allcolors=blue]{hyperref}
\usepackage{graphicx}
\usepackage[margin=1.3in]{geometry}% Make margins more narrow
\usepackage[numbers]{natbib}
\usepackage{notoccite}

\title{Pixelate to communicate: visualising uncertainty in maps of disease risk and other spatial continua}

\author{Aimee R Taylor$^{1,a,b}$, James A Watson$^{c,d}$, Caroline O Buckee$^a$}
\date{}

\begin{document}
\maketitle

\noindent \scriptsize {\textsuperscript{1}Corresponding author: ataylor@hsph.harvard.edu} \\

\noindent  \scriptsize {$^a$Department of Epidemiology, Harvard T. H. Chan School of Public Health, Boston, MA 02115, USA; $^b$Broad Institute of MIT and Havard, Cambridge, MA, 02142, USA; $^c$Mahidol-Oxford Tropical Medicine Research Unit, Faculty of Tropical Medicine, Mahidol University, Bangkok, 10400, Thailand; $d$Centre for Tropical Medicine and Global Health, Nuffield Department of Medicine, University of Oxford, New Richards Building, Old Road Campus, Roosevelt Drive, Oxford, OX3 7LG, UK} \\

\noindent \small{ \textbf{Keywords:} Uncertainty visualisation; geostatistics; isopleth risk map; \textit{Plasmodium falciparum} 

\section*{Abstract}
Maps have long been been used to visualise estimates of spatial variables, in particular disease burden and risk. Predictions made using a geostatistical model have uncertainty that typically varies spatially. However, this uncertainty is difficult to map with the estimate itself and is often not included as a result, thereby generating a potentially misleading sense of certainty about disease burden or other important variables. To remedy this, we propose simultaneously visualising predictions and their associated uncertainty within a single map by varying pixel size. We illustrate our approach using examples of malaria incidence, but the method could be applied to predictions of any spatial continua with associated uncertainty. 

\section*{Introduction}
% Why this now: disease mapping
The contemporary relevance of disease mapping cannot be overstated in view of pandemics such as covid-19. Since the removal of a contaminated water pump handle following John Snow's pioneering map of cholera in London (1854) \cite{tufte2001}, disease maps have informed decisions in public health and epidemiology. 
% Uncertainty communication in general
Increasingly, researchers, governments, and organisations like the WHO and the UN set their priorities based on spatially varying predictions of disease burden. Interpolation and extrapolation from observed data using statistical models allow for prediction of disease burden at arbitrarily fine scales. These predictions have associated uncertainty. Communication of spatially heterogeneous uncertainty is vital to avoid misinterpretation and to highlight data gaps. 

% Disease risk maps: clarify cloropleth vs isopleth
Maps of disease risk typically depict spatially varying predictions either shaded-in-area (cloropleth risk map) or distributed over a spatial continuum (isopleth risk map) \cite{tufte2001, goovaerts2006A2PKriging}. We focus on the latter. 
% Isopleths: deterministic vs stochastic
Isopleth maps can be used to depict spatial continua of any type, including fixed phenomena (e.g. elevation). 
% Isopleths: clarify contour lines vs pixels on grid
Traditional isopleth maps feature isopleths (e.g. contour lines). Contemporary ones tend to represent variations across spatial continua using colour and shade.

% Geostats 
Predictions depicted by isopleth maps are typically generated using geostatistical or geospatial models \cite{diggle1998, hengl2018}. %kelsall2002
(Geospatial is a general term that applies to a wide class of models including geostatistical models that are specifically concerned with spatial continua). To understand the uncertainty associated with predictions made using geostatistical models (and their geospatial counterparts), we briefly describe the geostatistical framework, albeit separate from that of visualisation.

% History
% Geostatistical models have long been used to model disease risk \cite{diggle1998}, %kelsall2002
% but they actually originate in the mining industry of the mid-twentieth century \cite{krige1951, matheron1963}. 
% krige1951 (the "original" geostats paper)
% matheron1963 (has a historical into)
% Since mining, applications progressed in fields of remote sensing, social science, and medical geography \cite{goovaerts2006A2PKriging}
% Early examples of visualisations of geostat. mapping of disease risk inc. childhood cancer (oliver1998), childhood malaria (diggle2002) and disease risk more generally (kelsall2002) 
% Essential geostats: 
Under a geostatistical model it is assumed that there is a latent spatially continuous stochastic process (e.g. unobserved malaria risk) that underpins a measurement variable whose realisations we can observe (e.g. malaria cases) \cite{diggle1998}. Given possibly noisy observations of the measurement variable at discrete sampling locations, a common goal is to predict this variable elsewhere (e.g. predict malaria cases at unsampled locations). Via the covariance structure of the stochastic process, the model imposes spatial correlation between observations. Typically, the predictive variance for locations distant from sampled locations exceeds that of nearby predictions. Sampling locations % which can be generated deterministically or stochastically (under a valid model the stochastic process on which the sampling locations depend must be independent from the one on which the measurement variable depends)
are often spatially heterogeneous (malaria case observations, for example, are almost always sparse and clustered in places with better access to care). As such, uncertainty can be highly variable in space (spatial heteroscedasticity). In addition to data on the measurement variable, spatial data on explanatory variables (e.g. elevation, vegetation) are sometimes included in the model to improve predictive accuracy and can thus influence the spatial heteroscedasticity.

% Back to timeliness
Advances in computational methods and large scale data collection, especially for explanatory variables, support highly resolved predictions (e.g. on a 5 km grid in the latest global malaria maps \cite{battle2019, weiss2019}). This has led to prolific high-profile mapping exercises in many fields, notably in public health, e.g. \cite{CGF0mapping, deribe2019, battle2019, weiss2019}. 
% Gap
Highly resolved maps are aesthetically pleasing. However, this high resolution can be extremely misleading, creating an illusion of precision where there is sometimes none. Maps of spatially varying uncertainty sometimes accompany predictions in the main text, e.g. \cite{battle2019}. 
However, due to restricted space, they are often incomplete (accompany a subset of predictions) or relegated to supplementary files. Other times they are simply ignored due to the difficulty of visual comparison across multiple maps; see illustrative example. 

% Response
We propose varying pixel size as a simple, visually intuitive method to merge predictions and their uncertainty, thereby ensuring uncertainty visualisation in a single map. To support our proposal and demonstrate our method, we provide a simple R package, \textbf{pixelate} (\href{https://github.com/artaylor85/pixelate}{github.com/artaylor85/pixelate}).

\section*{Results}

\subsection*{Methodological overview}
% Summary of concept and method (a result since contribution is methodological)
Our main result is a proposed method to visualise uncertainty in maps of disease risk and other spatial continua using pixelation. 
% concept: 
Specifically, we propose varying pixel size such that areas of high average uncertainty are unresolved, while areas with high average certainty are resolved (analogous to highly versus lowly resolved satellite images), thereby inviting a sense of precision only in areas where confidence is merited. 
% method overview: 
To vary pixel sizes, predictions are first grouped into a number of large initial pixels (squares or rectangles comprising many predictions), whose lower bound is specified. The average uncertainty of the predictions within each large initial pixel is then computed; and, for each large initial pixel, predictions are either averaged across it (if the average uncertainty within it is high), or across smaller pixels nested within it (if the average uncertainty within it is lower). The resulting plot of averaged predictions is deliberately and selectively pixelated, similar to a photo that is deliberately and selectively pixelated to disguise a person's identity.
% method specifics:
For each large initial pixel, the size of any nested pixels depends on the quantile interval into which the associated average uncertainty falls, where quantiles are based on the empirical distribution of average uncertainty across all large initial pixels, and the number of quantile intervals (thus different pixel sizes) is user-specified. Quantile interval allocation plots show how uncertainty varies across large initial pixels; while tables summarising pixel dimensions quantify how pixel size translates to average uncertainty (see the vignette of package \textbf{pixelate}). The smallest pixel contains a single prediction; the per-pixel prediction count of larger pixels is calculated according to above mentioned parameters (plus two others that control the rate at which nested pixels scale); see the documentation of the \texttt{pixelate} function in the \textbf{pixelate} package. Pixelation can thus be fine tuned according to the needs of the researcher. Pixelation parameters should be reported alongside pixelated maps, similar to model parameters. %e.g. regression coefficients and variogram parameters \cite{hengl2018}.  
It is also important to report if spatial covariance is accounted for, since pixelation relies on averaging uncertainty at the largest pixel size; see \cite{gething:10} and the vignette of the \textbf{pixelate} package for full details. 

% trim={<left> <lower> <right> <upper>}
\begin{figure}[H]
    %======================
    \begin{subfigure}{\columnwidth}
    \caption{Pixelated prediction map.}
    \includegraphics[trim={1cm 0 1cm 0cm},clip, page = 3, width=\columnwidth]{pixelate_plots}
    \vspace{-0.2cm}
    \label{sfig:pix}
    \end{subfigure}
    %======================
    \begin{subfigure}{0.5\columnwidth}
    \caption{Unpixelated prediction map}
    \includegraphics[trim={1cm 0 1cm 0cm}, clip, page = 1, width=\columnwidth]{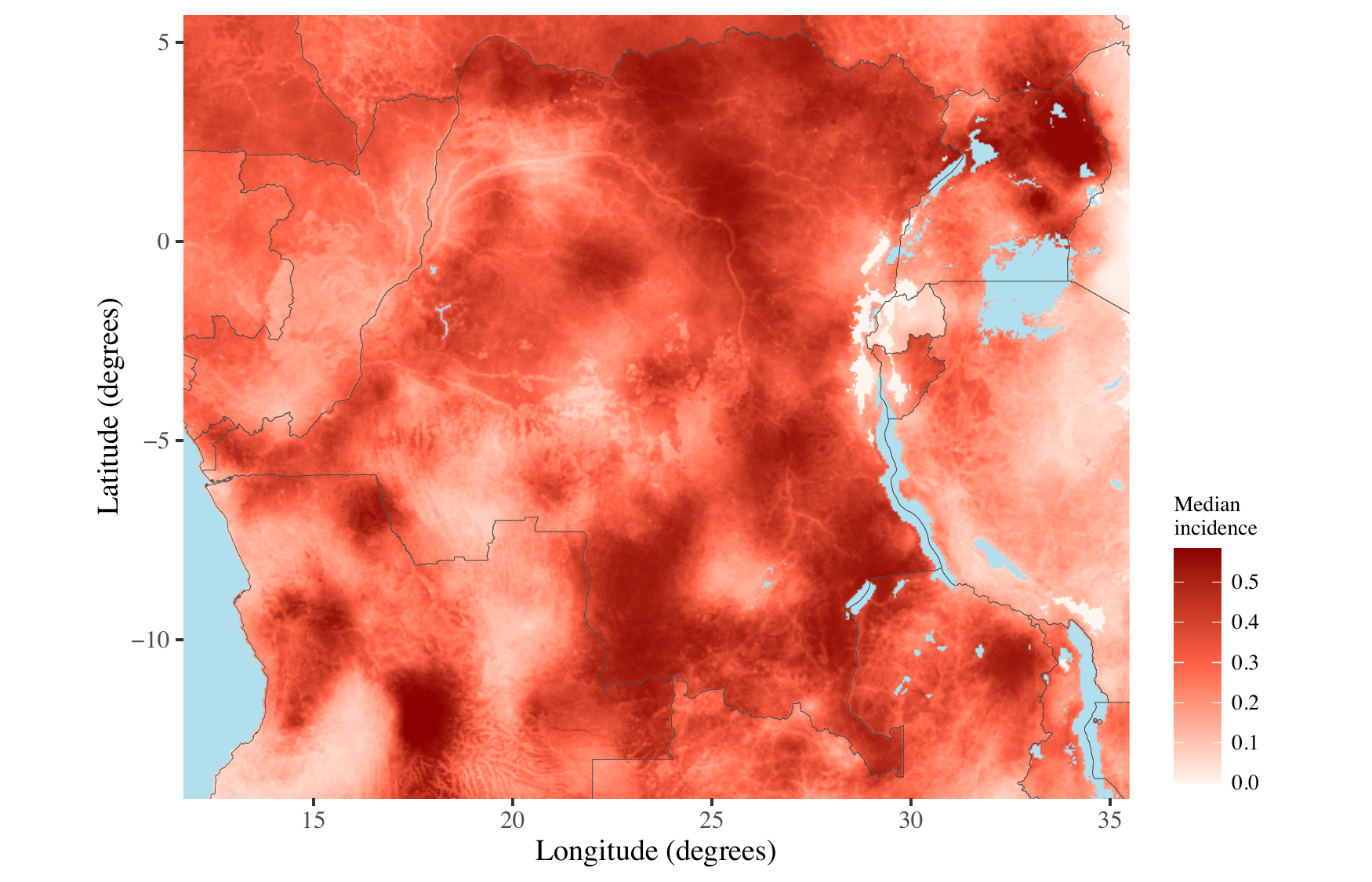}
    %---------------------------
    \begin{picture}(0,0) 
    % coordinates from bottom left
    \put(24,30){\includegraphics[height = 1.5cm, page = 4]{pixelate_plots}}
    \end{picture}
    %---------------------------
    \vspace{-0.2cm}
    \label{sfig:pred}
    \end{subfigure}
    %======================
    \begin{subfigure}{0.5\columnwidth}
    \caption{Uncertainty map}
    \includegraphics[trim={1cm 0 1cm 0cm}, clip, page = 2, width=\columnwidth]{pixelate_plots}
    \vspace{-0.5cm}
    \label{sfig:uncert}
    \end{subfigure}
    %======================
\caption{\small Pixelation of predicted 2017 \textit{P. falciparum} incidence (cases per person per annum) in central Africa \cite{weiss2019}. Panel \ref{sfig:pix} shows pixelated predictions; panels \ref{sfig:pred} and \ref{sfig:uncert} show the original predictions and their uncertainties (adaptations of Figures 4 and S40 of \cite{weiss2019}). The bottom-left inset of panel \ref{sfig:pred} delineates the central African region. The lower bound on the number of large initial pixels in both the horizontal and vertical direction was set to 12. Other parameters were set to default values: six different pixel sizes (zoom to see the smallest) that scale by iterative multiplication with factor one.} 
\end{figure}

\subsection*{Mapping malaria burden with uncertainty}
Our proposed methodological approach supports spatially continuous predictions of any kind provided that they have an associated measure of uncertainty. We illustrate our method using predicted malaria incidence: in a single map (Figure \ref{sfig:pix}) we represent both predictions of \textit{Plasmodium falciparum} incidence and the corresponding uncertainty (Figures \ref{sfig:pred} and \ref{sfig:uncert}, adaptations of Figures 4 and S40 of \cite{weiss2019}). % In this example, a prediction is a posterior predictive median value, while uncertainty is the range from the 0.250 to 0.975 quantile of the posterior predictive distribution. Alternatively, we could have taken other measures of location and uncertainty, e.g. the mean and standard deviation as in Van Den Hoogen et al. Nature 2019. 

In Figure \ref{sfig:pix}, areas with high average uncertainty around the predicted \textit{P. falciparum} incidence are visualised with large pixels and are thus unresolved, e.g. parts of the Democratic Republic of the Congo; areas of relative certainty are visualised with smaller pixels and are thus more resolved, e.g. Uganda. 
% Note on blue and white regions
The pixelated risk map also includes areas of missing predictions (blue) and of predictions that are zero with certainty (white). These are excluded from the pixelation process and thus appear exactly as they do in the original unpixelated map of predicted incidence (Figure \ref{sfig:pred}). 

% Decision making 
By merging into a single map (Figure \ref{sfig:pix}) predictions (Figure \ref{sfig:pred}) and their uncertainty (Figure \ref{sfig:uncert}), pixelation allows the consumer of the map to make decisions based on both types information concurrently. For example, considering enhanced surveillance, areas of high and low priority can be identified rapidly: high priority areas with high but uncertain risk are pixelated and dark; low priority areas with low but uncertain risk are pixelated and pale (Figure \ref{sfig:pix}). Meanwhile, resources can be allocated rapidly among areas with adequate surveillance: high priority areas with certain high risk are resolved and dark; low priority areas with certain low risk are resolved and pale (Figure \ref{sfig:pix}). Such operationally relevant information - critical for policy markers - cannot be extracted from the standalone map of unpixelated predictions (Figure \ref{sfig:pred}), and is difficult to extract rapidly from Figures \ref{sfig:pred} and \ref{sfig:uncert} side-by-side: without referring to Figure \ref{sfig:pix}, we challenge the reader to identify accurately and rapidly areas of 1) certain high risk; 2) uncertain high risk; 3) uncertain low risk and 4) certain low risk. 

% Sources of certainty. This section is hard to write conclusively because the meaning of API "input data" in the supplementary file of Weiss2019 is unclear. E.g. Rwanda appears to have little epidemiological data based on Figure S1 of Weiss2019 (between 2000 and 2016, only 2-20 observations in 2012), yet a lot of API "input data" (which is not defined) based on Figures S5-S8 of Weiss2019.
The variation in pixel size in Figure \ref{sfig:pix} invites a sense of precision only in areas where confidence is merited. Among these resolved regions, 
high certainty is expected where epidemiological data are dense, e.g. likely Uganda (assuming that the temporal density shown in Figure S1 of \cite{weiss2019} correlates with spatial density). However, information from explanatory variables can also contribute to increased certainty.
%Termed hybrid spatial pattern in Hengl et al. 2018. 
For example, despite presumably sparse epidemiological data, the central region of the Albertine rift has low uncertainty (including many predictions that are zero with certainty), likely due to elevation. %(we cannot say for sure since there are not maps of explanatory variables in the supplementary appendix of \cite{weiss2019}). 
In future work, we will develop interactive maps that provide per-pixel summaries of the different contributions to average certainty. However, this requires enhancing model output as well as its visualisation.

%===============================
\section*{Discussion}
%===============================

% Comparison to existing literature 
The problem of visualising uncertainty in maps of spatial continua is notoriously hard. We propose one solution based on pixelation. Pixelation has been used previously to communicate uncertainty in chloropleth maps \cite{lucchesi2017}. The R package \textbf{Vizumap} (\href{https://github.com/lydialucchesi/Vizumap}{github.com/lydialucchesi/Vizumap}) includes a function \texttt{pixelate} for cloropleth map pixelation. It also supports other visualisations of mapped uncertainty including bivariate choropleth and exceedance probability (EP) maps \cite{lucchesi2017, kuhnert2018}. Bivariate maps (e.g. of disease prevalence and its uncertainty \cite{CGF0mapping}) have been described as visual puzzles: very sophisticated but not very intuitive \cite{tufte2001}. EP maps convey probabilities (e.g. the probability that a disease exceeds a specified prevalence \cite{deribe2019}), % Another example: Nature 123
thus provide actionable insight for policy makers \cite{kuhnert2018}. They do not circumvent an illusion of precision, however.  

% Limitations
Our approach averts misleading illusions of precision and is intuitive, but it does have it own limitations. The perception of area (e.g. pixel size) varies across different people \cite{tufte2001}. %p. 66. 
However, in most cases, mapping aims to provide a relative overview (e.g. to enable policy makers to identify quickly priority regions) not absolute numbers. Perception may also differ according to expertise: to a geostatistician, smoothness in the covariance structure is a hallmark of data sparsity, whereas we rely on a non-technical interpretation of smoothness: regions with relative certainty are depicted using smaller pixels and are thus more resolved, akin to an information rich satellite image. Since mapping is intended to transfer knowledge e.g. from the geostatistician to a policy maker, we think its more appropriate to rely on a non-technical interpretation. Other limitations are surmountable: if there is no spatial variation among pixelated predictions, spatial variation in uncertainty will be invisible. In this case, it would make more sense to plot uncertainty directly (e.g. Figure \ref{sfig:uncert}). Finally, it is important to note that pixelation is a visualisation approach only. Visualisation alone cannot correct for inadequate model output (e.g. if the coverage of the prediction variance is poor, if the predictions are outdated, or if spatial covariance is unaccounted for).  

\section*{Conclusion}
Pixelation provides a simple, flexible, and intuitive way to combine spatially continuous predictions with their uncertainty thereby enhancing communication and veracity. Uncertainty visualisation for maps of spatial continua is especially important in public health, where increasingly maps of disease risk are central for policy planning and research. Our proposed method provides proof of concept. Experiments are needed to determine the impact of parameter choices, to check for visual distortion, and to compare with alternative approaches. 

\section*{Methods}
All code was written in R \cite{R}. % using the R packages \textbf{devtools} \cite{devtools} and \textbf{roxygen2} \cite{roxygen2} to develop the package \textbf{pixelate}. 
The package \textbf{pixelate} centres around a single function whose output is visualised using functions from the packages \textbf{ggplot2} \cite{ggplot2} and \textbf{sf} \cite{sf}. Shape file data used in the plots were obtained using function \texttt{getShp} from the R package \textbf{malariaAtlas} \cite{malariaAtlas}. Dark red was chosen to depict higher levels of both incidence and uncertainty following \cite{tufte2001}. The optimal choice and interpretation of colour is beyond the scope of this brief report, however. The script written to generate the plots, pixelate\_plots.R, is available online: \href{https://github.com/jwatowatson/Pixelation/blob/master/Code/pixelate\_plots.R}{github.com/jwatowatson/Pixelation}. 

% "Mind's eye does not readily give a visual ordering to colors, except possibly for red to reflect higher levels than other colors" meanwhile varying shades of grey do have natural ordering - [p154 Tuft]

We illustrate the our proposal using spatial predictions of \textit{P. falciparum} malaria incidence in 2017 \cite{weiss2019}, available from the Malaria Atlas Project. Specifically, we downloaded posterior predictive summaries of \textit{P. falciparum} incidence by selecting ANNUAL MEAN OF PF INCIDENCE at \href{https://malariaatlas.org/malaria-burden-data-download/}{map.ox.ac.uk/malaria-burden-data-download/}. We formatted the downloads using the script format\_pf\_incidence.R in the data-raw/ directory of the \textbf{pixelate} source package, available online (\href{https://github.com/artaylor85/pixelate}{github.com/artaylor85/pixelate}). 

\small {\section*{Acknowledgements}
A.R.T. and C.O.B. are supported by a Maximizing Investigators Research Award for Early Stage Investigators (R35 GM-124715). The funding source had no involvement in any part of the paper or the decision to submit it for publication. Thanks are extended to Luke Bornn for helpful discussion.} \\

\noindent \scriptsize{Conception and design: A.R.T; acquisition of data: J.A.W; supervision: C.O.B; interpretation and writing: all authors contributed. All authors have no conflict of interest to disclose.} \\

\normalsize
\bibliographystyle{unsrt}
\bibliography{ms}

\begin{thebibliography}{10}

\bibitem{tufte2001}
Edward~R Tufte.
\newblock {\em The visual display of quantitative information}, volume~2.
\newblock Graphics press Cheshire, CT, 2001.

\bibitem{goovaerts2006A2PKriging}
Pierre Goovaerts.
\newblock Geostatistical analysis of disease data: accounting for spatial
  support and population density in the isopleth mapping of cancer mortality
  risk using area-to-point poisson kriging.
\newblock {\em International Journal of Health Geographics}, 5(1):52, 2006.

\bibitem{diggle1998}
Peter~J Diggle, Jonathan~A Tawn, and RA~Moyeed.
\newblock Model-based geostatistics.
\newblock {\em Journal of the Royal Statistical Society: Series C (Applied
  Statistics)}, 47(3):299--350, 1998.

\bibitem{hengl2018}
Tomislav Hengl, Madlene Nussbaum, Marvin~N Wright, Gerard~BM Heuvelink, and
  Benedikt Gr{\"a}ler.
\newblock Random forest as a generic framework for predictive modeling of
  spatial and spatio-temporal variables.
\newblock {\em PeerJ}, 6:e5518, 2018.

\bibitem{battle2019}
Katherine~E Battle, Tim~CD Lucas, Michele Nguyen, Rosalind~E Howes, Anita~K
  Nandi, Katherine~A Twohig, Daniel~A Pfeffer, Ewan Cameron, Puja~C Rao, Daniel
  Casey, et~al.
\newblock Mapping the global endemicity and clinical burden of plasmodium
  vivax, 2000--17: a spatial and temporal modelling study.
\newblock {\em The Lancet}, 394(10195):332--343, 2019.

\bibitem{weiss2019}
Daniel~J Weiss, Tim~CD Lucas, Michele Nguyen, Anita~K Nandi, Donal Bisanzio,
  Katherine~E Battle, Ewan Cameron, Katherine~A Twohig, Daniel~A Pfeffer,
  Jennifer~A Rozier, et~al.
\newblock {Mapping the global prevalence, incidence, and mortality of
  Plasmodium falciparum, 2000--17: a spatial and temporal modelling study}.
\newblock {\em The Lancet}, 394(10195):322--331, 2019.

\bibitem{CGF0mapping}
{Local Burden of Disease Child Growth Failure Collaborators and others}.
\newblock Mapping child growth failure across low-and middle-income countries.
\newblock {\em Nature}, 577(7789):231, 2020.

\bibitem{deribe2019}
Kebede Deribe, Aimable Mbituyumuremyi, Jorge Cano, Mbonigaba~Jean Bosco,
  Emanuele Giorgi, Eugene Ruberanziza, Ursin Bayisenge, Uwayezu Leonard,
  Jean~Paul Bikorimana, Aniceth Rucogoza, et~al.
\newblock Geographical distribution and prevalence of podoconiosis in rwanda: a
  cross-sectional country-wide survey.
\newblock {\em The Lancet Global Health}, 7(5):e671--e680, 2019.

\bibitem{gething:10}
Peter~W Gething, Anand~P Patil, and Simon~I Hay.
\newblock {Quantifying aggregated uncertainty in Plasmodium falciparum malaria
  prevalence and populations at risk via efficient space-time geostatistical
  joint simulation}.
\newblock {\em PLoS Computational Biology}, 6(4):e1000724, 2010.

\bibitem{lucchesi2017}
Lydia~R Lucchesi and Christopher~K Wikle.
\newblock Visualizing uncertainty in areal data with bivariate choropleth maps,
  map pixelation and glyph rotation.
\newblock {\em Stat}, 6(1):292--302, 2017.

\bibitem{kuhnert2018}
PM~Kuhnert, DE~Pagendam, R~Bartley, DW~Gladish, SE~Lewis, and ZT~Bainbridge.
\newblock Making management decisions in the face of uncertainty: a case study
  using the burdekin catchment in the great barrier reef.
\newblock {\em Marine and Freshwater Research}, 69(8):1187--1200, 2018.

\bibitem{R}
{R Core Team}.
\newblock {\em R: A Language and Environment for Statistical Computing}.
\newblock R Foundation for Statistical Computing, Vienna, Austria, 2018.

\bibitem{ggplot2}
Hadley Wickham.
\newblock {\em ggplot2: Elegant Graphics for Data Analysis}.
\newblock Springer-Verlag New York, 2016.

\bibitem{sf}
Edzer Pebesma.
\newblock {Simple Features for R: Standardized Support for Spatial Vector
  Data}.
\newblock {\em {The R Journal}}, 10(1):439--446, 2018.

\bibitem{malariaAtlas}
Daniel Pfeffer, Tim Lucas, Daniel May, Joseph Harris, Jennifer Rozier,
  Katherine Twohig, Ursula Dalrymple, Carlos Guerra, Catherine Moyes, Mike
  Thorn, Michele Nguyen, Samir Bhatt, Ewan Cameron, Daniel Weiss, Rosalind
  Howes, Katherine Battle, Harry Gibson, and Peter Gething.
\newblock malariaatlas: an r interface to global malariometric data hosted by
  the malaria atlas project.
\newblock {\em Malaria Journal}, 17(1):352, 2018.

\end{thebibliography}

\end{document}